\documentclass[twocolumn,showpacs,preprintnumbers,amsmath,amssymb]{revtex4}
\usepackage{graphicx}% Include figure files
\usepackage{dcolumn}% Align table columns on decimal point
\usepackage{bm}% bold math

\begin{document}

%\preprint{APS/123-QED}

\title{Can experiments select the configurational component of excess entropy?}

\author{
    Silvia Corezzi, Lucia Comez, and Daniele Fioretto}
%\email{Lucia.Comez@fisica.unipg.it}
\address{
     $^1$
     Dipartimento di Fisica and INFM, Universit\`a di Perugia,
     I-06100, Perugia, Italy.}

\date{November 15, 2002}

\begin{abstract}

We introduce an experimental method of assessing the vibrational
and configurational components of the excess entropy of a liquid
over crystal, based on a joined investigation of dynamic and
thermodynamic properties as a function of temperature and
pressure. We analyze light scattering, dielectric, calorimetric
and dilatometric measurements of three prototype glass formers,
orthoterphenyl, salol, and glycerol. In all cases we find that
about 70\% of the excess entropy is configurational in nature.

\end{abstract}

%\pacs{64.70.Pf, 64.90.+b, 78.35.+c}

\maketitle

Liquids become glasses as a consequence of the suppression of
long range molecular mobility. The most evident feature that
accompanies glass formation is a dramatic increase of the
structural relaxation time, $\tau_\alpha$. A key for
understanding this phenomenon is the Adam-Gibbs (AG) theory
\cite{AG}, which states a relationship between $\tau_\alpha$ and
the configurational entropy $S_{c}$,
\begin{equation}
  \tau=\tau_0 \exp \left( \frac{C}{T S_c} \right).
  \label{AG}
\end{equation}
Here, $S_{c}\equiv S^{melt}-S^{vibr}$ where $S^{melt}$ is the
entropy of the metastable liquid phase, and $S^{vibr}$ is the
entropy of an ideal amorphous-solid phase (ideal glass) in which
only vibrations are active; $C$ is nearly constant, and $\tau_0$
is the relaxation time in the high temperature limit.

The difficult point in testing the predictions --- and the
validity itself --- of Eq.~(\ref{AG}) is the evaluation of the
configurational entropy from experimental data. In the past, this
problem has been overcome by approximating the vibrational heat
capacity at low temperatures by the heat capacity of an
experimentally accessible solid phase, i.e., the crystal or the
glass \cite{YamTsu98}. Therefore, \emph{configurational} entropy
is replaced by \emph{excess} entropy. Notwithstanding, the AG
equation succeeds in describing experimental results. This fact
looks like a paradox, since the same evidence is also found in
computer simulation studies, where the actual configurational
entropy is calculated. An approximate proportionality between
configurational and total excess entropies could be a possible
explanation of the paradox. From Eq.~(\ref{AG}) it is evident that
relaxation time measurements performed as a function of
temperature alone, cannot help to distinguish between $S_c$ and a
quantity proportional to it, as the proportionality constant
simply renormalizes the value of $C$.

Numerical simulations of supercooled water, and analytical models
for thermodynamics of defect crystals (see
Ref.~\cite{AngellJNCS02}), seem to support this idea. On the other
hand, to our knowledge, the only experimental work assessing the
two separate contributions to excess entropy has been made on a
very simple material, elemental selenium \cite{Phillips89}. In
this case, the vibrational entropy over crystal is calculated
under harmonic approximation from the analysis of
neutron-scattering data, and it is subtracted from $S_{exc}$ to
determine $S_{c}$. This result gave rise to controversial
conclusions about the method of calculation of $S_{c}$, and the
existence of proportionality between $S_{c}$ and $S_{exc}$
(\cite{AngellJNCS02,JohariJNCS02}).

Here, we enter the debate trying to answer the question: ``Can
experiments select the configurational component of excess
entropy in glass forming systems?''. We show that calorimetric,
dilatometric, and relaxation measurements as a function of both
temperature $T$ and pressure $P$, allow us to extract the
configurational fraction of the excess entropy. Our method is
tested on three different prototype glass formers, orthoterphenyl,
salol, and glycerol and for relaxation data, obtained by
dielectric spectroscopy and light scattering.

The configurational entropy of a system at a given $T$ and $P$ can
be separated into an isobaric contribution at zero pressure, and
an isothermal contribution at temperature T \cite{Casalini}:

\begin{eqnarray}
 S_c(T,P)&&=S_{c}^{isob}(T,0)+S_c^{isoth}(T,P)= \nonumber\\
         &&=\int_{T_K}^T \frac{\Delta C_P(T')}{T'}dT'
-\int_0^P \Delta \left( \frac{\partial V}{\partial T} \right)
_{P'}dP',\nonumber\\ \label{TECE}
\end{eqnarray}
where $\Delta C_{P}=C_{P}^{melt}-C_{P}^{vibr}$ is the
configurational heat capacity at zero pressure,
$\Delta\left(\partial V/\partial T\right)_{P}=\left(\partial
V/\partial T\right)_{P}^{melt}-\left(\partial V/\partial
T\right)_{P}^{vibr}$ is the configurational thermal expansivity
at $T$, and $T_K$ is the temperature where the configurational
entropy of a liquid vanishes at $P=0$. Substituting Eq.
(\ref{TECE}) in Eq.~(\ref{AG}) gives the pressure extended AG
equation (PEAG) \cite{Casalini}. For evaluation purposes, also in
this approach one approximates the isobaric part of the
configurational entropy with the excess entropy, i.e.,
$S_{c}^{isob}(T,0)\approx S_{exc}(T,0)$. Within this
approximation, the PEAG model is able to describe experimental
data fairly well in the T-P space \cite{Casalini,ComezPRE02}. The
cost to be payed is a value of $\left(\partial V/\partial
T\right)_{P}^{vibr}$ which in some cases may considerably depart
from the expected value. This anomaly is a consequence of the poor
approximation $S_c \approx S_{exc}$. In contrast, in this work we
improve the PEAG model, we obtain a reasonable value for
$\left(\partial V/\partial T\right)_{P}^{vibr}$, and, more
interestingly, we determine a realistic estimate of the
configurational entropy.

\begin{figure*}
\scalebox{1.18}{
\includegraphics{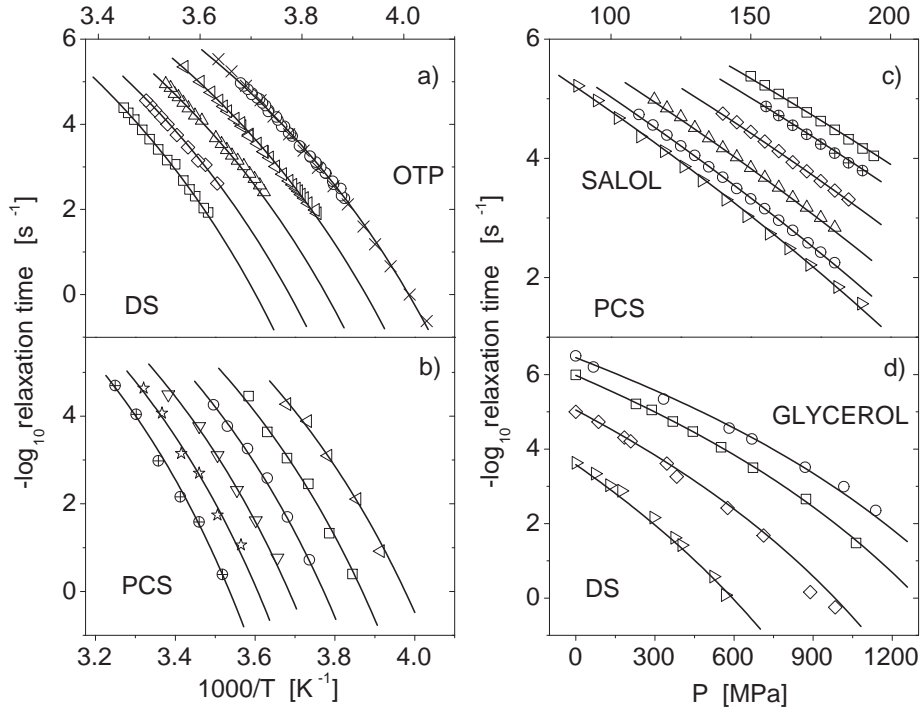}}
\caption{\label{tau} a) Structural relaxation time of OTP from
dielectric measurements at different pressures. Open symbols are
data taken  at different pressures \cite{NaokiEndouJPC87} ($P$=0.1
MPa ($\circ$), 19.6 MPa ($\vartriangleleft$), 39.2 MPa
($\vartriangle$), 58.8 MPa ($\lozenge$), 78.5 MPa ($\square$)).
Crosses are data at ambient pressure \cite{HansenStickelJCP97}.
 b) Structural
relaxation time $\langle\tau\rangle$ of OTP from
photon-correlation measurements at different pressures. Data
taken from Fytas et al.\cite{FytasJPC83} ($P$=0.1 MPa
($\vartriangleleft$), 25 MPa ($\square$), 50 MPa ($\circ$), 75
MPa ($\vartriangleright$), 100 MPa ($\star$), 125 MPa
($\oplus$)). c) Structural relaxation time $\langle\tau\rangle$
of salol from photon correlation measurements at different
temperatures \cite{ComezPRE02} ($T$=267.0 K
($\vartriangleright$), 268.5 K ($\circ$), 271.1 K
($\vartriangle$), 274.5 K ($\lozenge$), 278.2 K ($\oplus$), 280.4
K ($\square$)). d) Structural relaxation time of glycerol from
dielectric measurements at different temperatures (data taken
from Ref.~\onlinecite{JohariWhalley72}): $T$=217.5 K
($\vartriangleleft$), $T$=230.6 K ($\lozenge$), $T$=240.9 K
($\square$), $T$=247.3 K ($\circ$). In all panels, the solid
lines represent the best fit with the PEAG equation
Eq.~(\ref{AGphi}). The relaxation time for dielectric
measurements is $\tau=1/2\pi\nu_{max}$, with $\nu_{max}$ the
frequency of maximum dielectric loss. The average relaxation time
for photon correlation data is given by $\langle \tau
\rangle=(\tau_{K}/\beta_{K})\Gamma(\beta_{K}^{-1})$, where
$\tau_{K}$ and $\beta_{K}$ are the Kholrausch-Williams-Watts
\cite{KWW} characteristic time and stretching parameter,
respectively, and $\Gamma$ denotes the Euler gamma function.}
\end{figure*}

{\it The model}.--- Guided by the idea that an approximate
proportionality may exist between configurational and excess
entropies, we shall assume that the isobaric contribution to $S_c$
can be expressed, to first approximation, as a fraction of the
total excess entropy:
\begin{equation}
  S_c(T,0)=\Phi S_{exc}^{isob}(T,0).
  \label{Scphi}
\end{equation}
\noindent Hence, we modify the PEAG formula for the structural
relaxation time as follows:
\begin{equation}
  \tau(T,P)=\tau_0 \exp \left[ \frac{C}{T(\Phi S_{exc}^{isob}+S_c^{isoth})}
  \right];
  \label{AGphi}
\end{equation}
here, the parameter $\Phi$ ($\leq$1) quantifies the fraction of
excess entropy at atmospheric pressure configurational in nature.
Note that the presence of the term $S_c^{isoth}$ in
Eq.~(\ref{AGphi}) prevents the parameter $\Phi$ to play the role
of a simple renormalization constant.

The liquid over crystal excess entropy at atmospheric pressure,
$S_{exc}^{isob}(T,0)$, can be evaluated from calorimetric
measurements. In calculating the isothermal part of the
configurational entropy, $S_{c}^{isoth}(T,P)$, we do not make any
modification of the expression given in Ref.~\cite{Casalini},
since no mention to a reference state --- crystal or glass --- was
necessary. Accordingly, we approximate the vibrational thermal
expansivity at $P$ by its value at ambient pressure, i.e.,
$\left(\partial V/\partial T\right)_{P}^{vibr}\approx
\left(\partial V/\partial T\right)_{0}^{vibr}$
[Eq.~(\ref{TECE})]. This is due to the fact that the value and the
pressure dependence of the thermal expansivity for crystals and
glasses is usually much smaller than that for liquids. Therefore,
we are left with calculating the integral over pressure of the
thermal expansivity of the melt. Starting from the Tait equation
of state \cite{Tait}, simple algebra allows us to express this
contribution in terms of thermodynamic parameters obtained from
dilatometric measurements \cite{Casalini}.

{\it Testing the model}.--- A complete test of the model requires
the use of information determined by several experimental
techniques, i.e, $T$ and $P$-dependence structural relaxation
times, extended calorimetric data, pressure-volume-temperature
(PVT) measurements. In the following we consider three model
glass-formers: orthoterphenyl (OTP), phenyl salycilate (salol),
and glycerol. For these systems all the needed data are available,
and can be properly used to test the generality of our approach.
Altogether, calorimetric, volumetric and $\tau(T,0)$ data provide
all the parameters in Eq.~(\ref{AGphi}) --- for details see
Ref.~\cite{Casalini} --- except for $C$, $\Phi$ and
$\left(\frac{\partial V}{\partial T}\right)^{vibr}_{0}$, which
can be obtained by a multi-variable fit on experimental
relaxation data as a function of $T$ and $P$.

{\it Orthoterphenyl}.--- Isobaric heat capacity of crystalline,
supercooled and stable liquid OTP determined via calorimetry
\cite{ChangBestulJCP72} has been used to calculate the liquid
over crystal excess entropy $S_{exc}$ as a function of
temperature. \noindent Specific volume data of the liquid and
glassy OTP as a function of both temperature and pressure
\cite{NaokiKoedaJPC89} have been used to determine the parameters
of the Tait equation of state. Structural relaxation times
obtained by means of two different techniques are available from
the literature. A study of pressure effects on the dielectric
relaxation of supercooled OTP \cite{NaokiEndouJPC87} reports
several isobaric measurements as a function of temperature which
are shown in Fig.~\ref{tau}a together with dielectric data at
ambient pressure \cite{HansenStickelJCP97}. A second pressure and
temperature dependent photon correlation investigation has been
reported in \cite{FytasJPC83}. The average relaxation time
$\langle \tau \rangle$ appears in Fig.~\ref{tau}b in the $T$ and
$P$ ranges investigated.\\
We fit according to Eq.~(\ref{AGphi}) the data in the $T$ range
247-272 K at atmospheric pressure, where $\log_{10}\tau$ is
linear vs. $(T S_{exc})^{-1}$), and all the isobaric dielectric
data at higher pressures. The simultaneous fit gives:
$\Phi=0.69\pm 0.05$, $\left(\partial V/\partial T
\right)_{0}^{vibr}\!\!=(4.3\pm0.7)\times 10^{-8}$ \mbox{m$^{3}$
mol$^{-1}$ K$^{-1}$}, $C=(1.49\pm0.12)\times 10^{5}$ \mbox{J
mol$^{-1}$}. It is evident from Fig.~\ref{tau}a the high quality
of the agreement between experimental relaxation time data for
OTP (symbols) and calculated curves (solid lines) using the PEAG
model in the form of Eq.~(\ref{AGphi}). As a cross-check of the
reliability of the fit performed on the dielectric relaxation
data, we plot the \emph{same} curves as in Fig.~\ref{tau}b on the
PCS data. We find that the PCS data are perfectly described by the
same fitting parameters as the dielectric data, except for a
vertical shift corresponding to a different value of $\tau_0$
connected with the specific technique used.
%We observe that Eq.~(\ref{AGphi}) presents a
%great sensitivity in adjusting free parameters forcing the
%results to realistic values. Indeed,
It is worth stressing that the best-fit is set by a physical value
of the vibrational thermal expansivity: the obtained value
compares well with $\sim 5\times 10^{-8}$ \mbox{m$^{3}$
mol$^{-1}$ K$^{-1}$}, that is expected for an OTP-based solid form
of matter \cite{NaokiKoedaJPC89}. Reasonably, the value of the
vibrational thermal expansivity corresponds to about 30\% of the
melt thermal expansivity. We remark that using the PEAG form with
a preset $\Phi=1$ we find an unphysical value for the vibrational
thermal expansivity ($2.3\times 10^{-9}$ \mbox{m$^{3}$ mol$^{-1}$
K$^{-1}$}), about two order of magnitude lower than the
expansivity of the melt.

{\it Salol}.--- Precise determination of the heat capacity of
salol under atmospheric pressure is due to Oguni
\cite{OguniPrivate}.
%The temperature behavior of $S_{exc}$ is
%well represented by a linear function of $(1/T)$.
Dilatometric measurements $V(T,P)$ of crystalline and liquid
salol have recently been carried out \cite{ComezInpreparation}
permitting to extract the parameters of the Tait equation of
state. Isothermal relaxation times measured by photon-correlation
spectroscopy are available \cite{ComezPRE02}. The pressure
variation of $\langle\tau \rangle$ is shown in Fig.~\ref{tau}c in
the temperature range investigated. We analyze the relaxation
times data using Eq.~(\ref{AGphi}), and find $\Phi=0.68\pm0.08$,
$\left(\partial V/\partial T \right)_{0}^{vibr}=(3.8\pm
0.7)\times 10^{-8}$ \mbox{m$^{3}$ mol$^{-1}$ K$^{-1}$}, and
$C=(1.9\pm0.3)\times 10^{5}$ \mbox{J mol$^{-1}$}. The value for
the vibrational thermal expansivity is in remarkable agreement
with that found for the poly-crystal via PVT measurements
[$\left(\partial V/\partial T \right)_{0}^{crystal}\sim (4.0\pm
0.5)\times 10^{-8}$ \mbox{m$^{3}$ mol$^{-1}$ K$^{-1}$}]
\cite{ComezInpreparation}. A previous analysis of these data
performed by some of us \cite{ComezPRE02} using a preset
$\Phi=1$, i.e. replacing $S_c$ with $S_{exc}$, provided
$\left(\partial V/\partial T \right)_{0}^{vibr}=(1.0 \pm
0.7)\times 10^{-8}$ \mbox{m$^{3}$ mol$^{-1}$ K$^{-1}$}, which is
in feeble agreement with the experimental value of the
poly-crystal.

\begin{figure}
\scalebox{0.85}{
\includegraphics{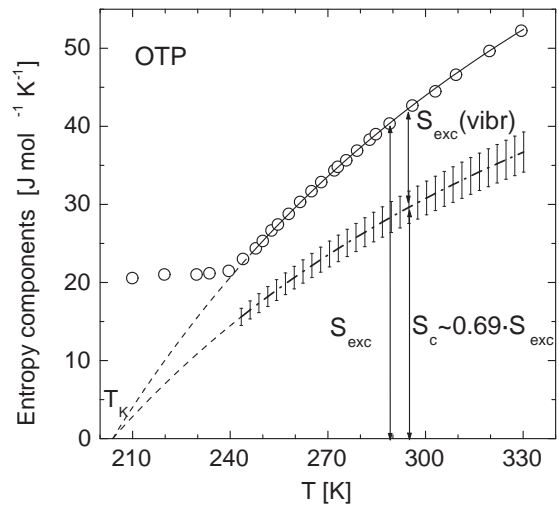}}% Here is how to import EPS art
\caption{\label{entropy} Entropy components in OTP at atmospheric
pressure. Circles represent the excess entropy of the liquid over
the crystal, $S_{exc}=S^{melt}-S^{crystal}$, calculated from
experimental calorimetric data. The  best fit with
$S_{exc}(T)=S_{\infty}-k/T$ for T$>$247 K, and its extrapolation
for T$<$247 K are shown with solid and dash line, respectively.
The thick dash-dot line represents the configurational entropy
evaluated as a fraction of the excess entropy, $S_{c}=\Phi
S_{exc}$. \mbox{$T_{K}$=203.8 K} is the temperature at which the
entropy of the liquid equals the entropy of the crystal.}
\end{figure}

{\it Glycerol}.--- Isobaric heat capacity of liquid-glass and
crystal glycerol \cite{GibsonGiauque23} have been used to evaluate
the liquid over crystal excess entropy.
%In this case, the
%temperature behavior of $S_{exc}(T)$ is well reproduced by a
%3$^{rd}$ order polynomial.
Volumetric measurements of glycerol reported in Ref.
\cite{Gilchrist57} as a function of both $T$ and $P$ have been
represented in terms of the
Tait equation. \\
As far as $\tau(T,P)$ data, we focus on the pioneering work by
Johari and Whalley \cite{JohariWhalley72}, which provides
relaxation time data in agreement with those recently reported by
Lunkenheimer at atmospheric pressure \cite{Lunkenheimer}. For our
test, we subject to the fit procedure data up to $\sim 1$ GPa (see
Fig.~\ref{tau}d) excluding all isotherms at temperature higher
than 247 K, in order to remain in the validity range of the PEAG
model. Because of the not very high quality of the data, and with
the background of the analysis of OTP and salol in place, we
undertake the fit procedure with $\left(\partial V/\partial T
\right)_{0}^{vibr}$ constrained to a physical value. Dilatometric
measurements of glycerol provide an average value
$(1.68\pm0.01)\times 10^{-8}$ for the thermal expansivity of the
glassy state in the temperature range investigated. As a result
of a global fit procedure we find $C=(12.8\pm0.4)\times 10^{5}$
\mbox{J mol$^{-1}$}, and $\Phi=0.71\pm0.01$ \cite{errorgly}, this
value being surprisingly close to that found in OTP and salol.

In summary, the fundamental information we gain from the above
analysis is that at atmospheric pressure, about 70\% of the
excess entropy is of configurational origin for each one of the
molecular systems here considered. These results astonishingly
agree with the value found for $S_c/S_{exc}$ (0.69) in selenium
\cite{AngellJNCS02}, where the configurational entropy was
calculated from neutron scattering data. In addition, we observe
that simulation studies on SPC-E water
\cite{Scal00,Mart01,Starr01} have shown $S_c$ and $S_{exc}$ to be
proportional, with a constant ratio of about 0.77
\cite{AngellJNCS02}.

In conclusion, we have introduced a general method to calculate
the configurational contribution to the excess entropy for
molecular liquids. The general validity of the method and the
reasonableness of the hypothesis made are tested on three model
glass formers. The results obtained include the following points:
{\it i}) Experiments performed as a function of $T$ and $P$ are
actually able to identify the configurational fraction of excess
entropy, giving a positive answer to our initial question; {\it
ii}) The value of the configurational fraction of excess entropy
reconciles the results obtained by previous laboratory and
computational tests; {\it iii}) Configurational entropy is
confirmed to be a key concept in controlling the slow dynamics of
glass forming systems.

This last evidence is also confirmed by the energy landscape
approach to the glass transition \cite{nature}, as highlighted by
recent simulation studies (see, among others,
Refs.~\cite{Scal00,Scior99,Sast00,BucAnd00,AngDil00,LaNave02}). We
believe that more efforts should be devoted in order to clarify
the connection among the potential energy landscape features (see,
for instance, the equation of state proposed in
Refs.~\cite{LaNave02,Debe99}) and experimental results in $T$ and
$P$ domains.

The authors would like to particularly thank Prof. M. Oguni for
shearing calorimetric data on salol prior to publication.


\begin{references}
\bibitem{errorgly} Note that the small uncertainty of the $\Phi$ value is due to a
reduced number of fitting parameters in this case.
\end{references}
\end{document}